# An Efficient Epileptic Seizure Detection Technique using Discrete Wavelet Transform and Machine Learning Classifiers

Rabel Guharoy[1], Nanda Dulal Jana[2] and Suparna Biswas[3]

[1,2] Department of CSE, National Institute of Technology, Durgapur
[3] Department of Dept. of ECE, GNIT, Kolkata

**Abstract.** This paper presents an epilepsy detection method based on discrete wavelet transform (DWT) with Machine learning classifiers. Here DWT has been used for feature extraction as it provides a better decomposition of the signals in different frequency bands. At first, DWT has been applied to the EEG signal to extract the detail and approximate coefficients or different sub-bands. After the extraction of the coefficients, principal component analysis (PCA) has been applied on different sub-bands and then a feature level fusion technique is used to extract the main features in low dimensional feature space. Three classifiers name: Support Vector Machine (SVM) classifier, K-Nearest-Neighbor (KNN) classifier, and Naive Bayes (NB) classifier have been used in the proposed work for classifying the EEG signals. The raised method is tested over Bonn databases and provides a maximum of 100% recognition accuracy for KNN, SVM, NB classifiers.

## 1. Introduction

Epileptic seizure is a neurophysiological disease which can be detected by analysing brain signals which is generated by brain neurons as it is a neurological disorder. This can be cured by applying EEG and ECoG sensible patterns should be recognized by machine learning and statistical feature. EEG records electrical activity of brain. EEG recording provides non-invasive, non-painful, efficient aid to search the dynamic brain response with complexity. The different application areas arise from diagnosis usually pay attention to the spectral things coming from EEG for the purpose of controlling these types of neuro-electrical performances which may be detected in EEG. Filter Bank Processing, average-frequencies and RMS Bandwidth Calculations, Least-Squares Support Vector Machine, Simulation performances and Discussion have to be done [1].

Different neuro-imaging technical methodologies can be found in the form of X ray, angiography, electroencephalogram (EEG), magneto encephalogram (MEEG) CT scan, ultrasound, single photon emission computed tomography (SPECT) functional magnetic resonance imaging (fMRI), positron emission tomography (PET) in order to analyse and infer electro physiological incident happening within case of human brains. Different methodologies have been adopted for the purpose of classification of such kind of signals utilizing different machine learning algorithm like neural networks, fuzzy inference system, wavelets, statistical methods and few others [2]. EEG is recorded since electrodes serialized in a especial sampling. A general standard called the International 10/20 systems is used here. Inexpensive methods and give a continuous record of brain activity with best comparison to millisecond solution. High temporal solution can achieve by this tool and for these causes the elaborated method discoveries of dynamic cognitive processes have been reported using EEG and ERP (Event Related Potentials) methods. Data set, pre-processing, Fourier analysis, Features Extraction are some ways to handle this [3]. The time domain, frequency domain, and wavelet-based feature extraction techniques for classification of EEG signals have been indicated [15–17].

Electroencephalogram (EEG) signal-processing technique in which Literature review, Clinical EEG data set, Statistical feature extraction using discrete wavelet transform (DWT), Principal component analysis (PCA), Linear discriminant analysis (LDA), Naive Bayes classifier are applied [4]. Sets of data explained in A – E. All the data sets contain hundred single-channels of signal EEG of the 24.5 S [6]. Thus, the sampling rate of knowledge has been calculated as 174.31s. The process of attaining the time series gains 0.6 Hz to 86 Hz. The methodology provides the first step exploration for low-pass filter as 40 Hz. After that both the sets A, B of EEG record makings has been extracts from the surfaces of five best components using openly kept eyes and closed eyes. Rest of the 2 sets evaluated 5 different patients in epileptogenic zone(D). Here, tall tent arrangements see confliction of hemisphere of brain(C). In E set contains an invasion phenomenon which is selected based on whole recording sites that providing corrective action [7]. The recording of whole data sets, A, B has been done like excess cranial and C, D, E like intracranial [8]. Curvelet Transform, B. PCA Technique are some techniques [5]. The feature selection methods result in various advantages for the minimization of the time required for training period, these result in faster working of training methods by reducing processed data, reducing the improper data quantity—such type of data permanently share error decisions; and increase the degree of accurate performance by cleaning data [20].

We can be found by analysing the brain signals created with brain neurons. Neurons are joined to each other in a hard way to connect with human organs and generate signals. The monitoring of brain signals is mainly done using Electroencephalogram (EEG) and Electrocorticography (ECoG) media [9]. Seizure Prediction, Feature Extraction is the method to calculate result and Epilepsy is detected. The raised Long Short-Term Memory (LSTM) classifier categorizes it is 3 types of signals with up to 96% accuracy. In binary classification same as discovery of inter-ictal or ictal only, its accuracy increases to 99%. The EEG signal is modelled as wide sense non-stationary any signal. Hurst Exponent and Auto regressive Moving Average (ARMA) features are removed from all signal. That way, two various configurations of LSTM architecture: single-layered memory units and double-layered memory units are same constructed [10].

we can detect epileptic seizure by analyzing brain signals which is generated by brain neurons as it is a neurological disorder. This can cure by applying EEG and ECoG. Sensible patterns should be recognized by machine learning and statistical feature [9].

Epileptic seizure diagnosis and treatment helped in development of automated seizure onset detection systems and determines best classifier having highest classification rates. By the use of neural network techniques, support vector machine and radial basis function machine learning techniques this can be simplified to aid neurophysiologists and the EEG data has been analyse for the purpose of detection of epileptic patterns [2].

Electroencephalogram (EEG) data is detailed analyzed and classified for detecting and predicting epileptic seizure. Discrete wavelet transforms (DWT), Principal component analysis (PCA) and Linear discriminant analysis (LDA) techniques have been used in EEG data patterns. These techniques using K-NN has been proved to be 98% to 100% accurate in result hence cure the patient [4].

The present work contains three primary sections that can be segmented as follows: Section ''Proposed Methods'' illustrates the detailing about the proposal of classification method. Next ''Data sets description'' illustrates the set of data utilized of us. In the end the entire job has been completed in section ''Statistical analysis.'' 2. Proposed Method Figure 1. Block diagram of EEG signal to feature classification.

2. **Proposed Method**

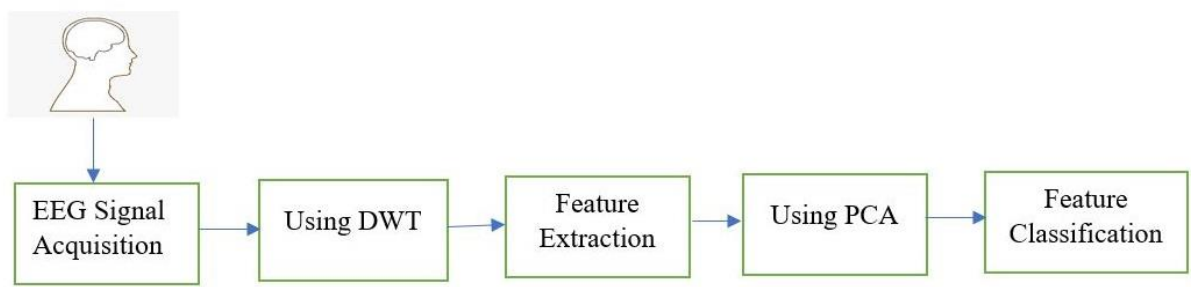

**Figure 1.** Block diagram of EEG signal to feature classification.

Using DWT to bring out the features as this is a good efficient process in comparison to the other processes in the view of accurate performance of extraction features in order to confirm the effective nature in case of the following moves, it was proved in several former studies [12][13]. The ICA algorithm can be found to be mostly used statistical technique that may be utilized in this area for signal noise removal, more specifically EEG signals. We study for this ICA based technique for the purpose of correcting and removing contaminations present in brain wave planning [18][19]. Mainly this technique is aimed at reducing the impact of surrounding parameters into EEG signals form.

Within work, we have suggested a different and novel proposal on epilepsy cases from EEG signals. Initially, we have taken some grade raw EEG signals those can be found in Bonn University where our way for 4 major steps: pre-processing, feature extraction, feature selection, and classification. The flow diagram of proposed method in shown is figure 1. Consists of total 4 module of feature classification, these are Discrete wavelet transform (DWT), feature extraction, using PCA and feature classification. We read the EEG signal then we have applied DWT. Extract features through used DWT, again used to 5 levels and the features as shows D1, D2, D3, D4, D5, and A5. We have used $5^{th}$ level of decomposition of figure 2.

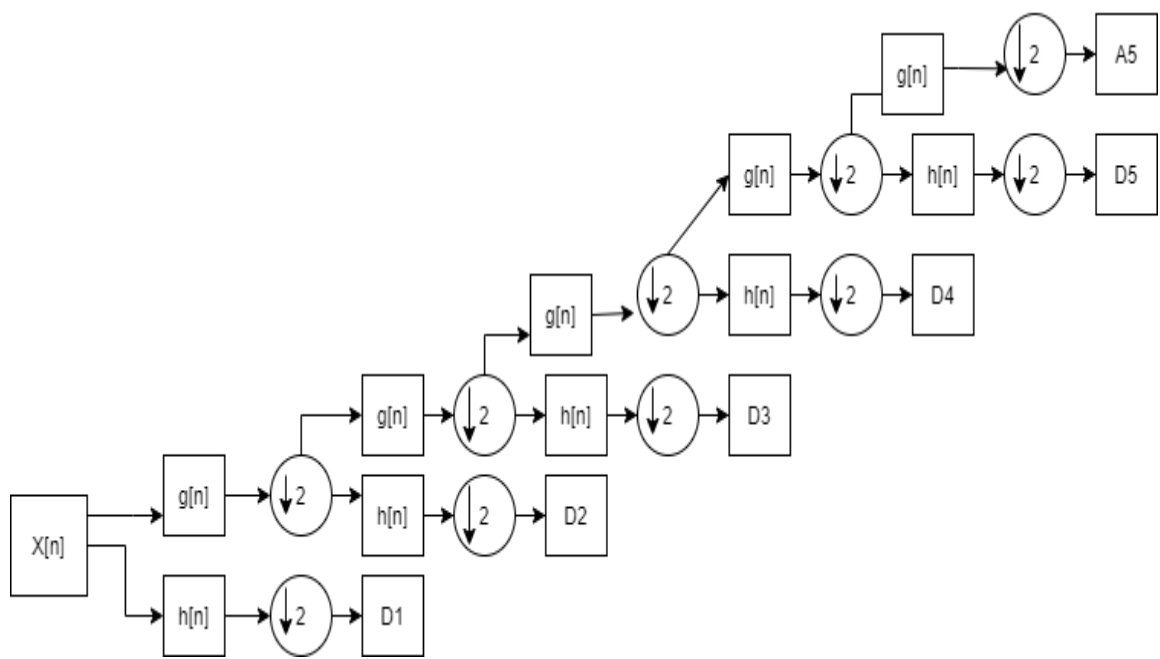

**Figure 2.** $5^{th}$ level of Decomposition.

Finally, we have got best result for $5^{th}$ level of decomposition. We have applied PCA to CA5, CD1, CD2, CD3, CD4 and CD5. We can use feature selection. Feature extraction and selection method explain in the figure 3. We are taken in six band ($CA_5$, $CD_1$, $CD_2$, $CD_3$, $CD_4$, $CD_5$) and applied PCA. We are extracted X no of principal component. We are taken as a 50 principal component as a X. After that we are used max function for five band then finally, we are applied classifier technique and final feature came.

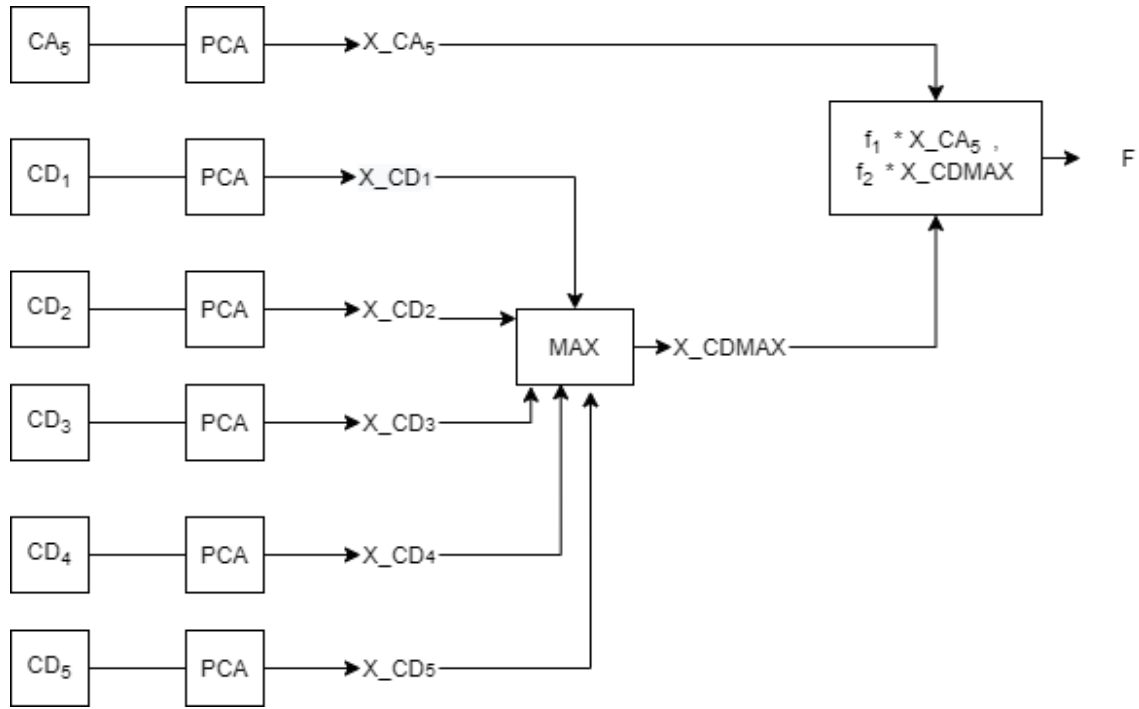

**Figure 3.** Block Diagram of feature extraction & selection.

We are using different type of classifiers, SVM, NB, KNN. A classifier is a technique it is utilize for different independent variable values (features) as input and calculates the similar class to which associated with independent variable [14].

3. **Result and discussion:**

Epilepsy data collected by the Department of Epilepsy at the University of Bonn, Germany, are gainable for gratis to public domain (Table 1).

| Set name | Annotation of data | Size | Acquisition circumstances |
| --- | --- | --- | --- |
| Set A | Z000.txt—Z100.txt | 564 KB | Five healthy subjects with open eye |
| Set B | O000.txt—O100.txt | 611 KB | Five healthy subjects with close eye |
| Set C | N000.txt—N100.txt | 560 KB | Five epileptics with seizure free status |
| Set D | F000.txt—F100.txt | 569 KB | Five epileptics with seizure-free status, inside five epileptogenic zone |
| Set E | S000.txt—S100.txt | 747 KB | Five subjects during seizure activity |

**Table 1.** Samples of data in normal and seizure cases.

The entire database is comprised of 5 separate sections of EEG signals, where such sections are symbolically denoted starting from A to E. Here, each section contains hundred signals, where time was recorded as 23.6 seconds. For the purpose of recording the data considering maximum accuracy, the authors utilized a system amplification of 128 signal channels, where the result outcome is found at sample rate 173.61 Hz and a depth of 12-bit resolution, counting the octave of 0.50–42 bandwidth. Considering the fact of the application of a band-pass filter (12 dB/Oct) [11].

| Case | Accuracy | Sensitivity | Specificity | Precision | Recall | F-measure |
|------|----------|-------------|-------------|-----------|--------|-----------|
| **A-C** | 97.500000 | 95.833333 | 100.000000 | 100.000000 | 95.833333 | 0.978723 |
| A-D | 95.000000 | 100.000000 | 94.736842 | 94.736842 | 100.000000 | 0.952381 |
| A-E | 100.000000 | 100.000000 | 100.000000 | 100.000000 | 100.000000 | 1.000000 |
| B-C | 97.500000 | 100.000000 | 96.000000 | 94.736842 | 100.000000 | 0.967742 |
| B-D | 92.500000 | 100.000000 | 94.444444 | 95.238095 | 100.000000 | 0.930233 |
| B-E | 100.000000 | 100.000000 | 100.000000 | 100.000000 | 100.000000 | 1.000000 |

**Table 2.** Results of proposed model with KNN Algorithm.

| Case | Accuracy | Sensitivity | Specificity | Precision | Recall | F-measure |
|------|----------|-------------|-------------|-----------|--------|-----------|
| **A-C** | 92.500000 | 100.000000 | 90.476190 | 90.000000 | 100.000000 | 0.923077 |
| A-D | 92.500000 | 95.238095 | 94.444444 | 95.000000 | 95.238095 | 0.930233 |
| A-E | 100.000000 | 100.000000 | 100.000000 | 100.000000 | 100.000000 | 1.000000 |
| B-C | 87.500000 | 91.304348 | 100.000000 | 100.000000 | 91.304348 | 0.893617 |
| B-D | 97.500000 | 96.000000 | 100.000000 | 100.000000 | 96.000000 | 0.979592 |
| B-E | 100.000000 | 100.000000 | 100.000000 | 100.000000 | 100.000000 | 1.000000 |

**Table 3.** Results of proposed model with SVM Algorithm.

| Case | Accuracy | Sensitivity | Specificity | Precision | Recall | F-measure |
|------|----------|-------------|-------------|-----------|--------|-----------|
| **A-C** | 87.500000 | 100.000000 | 82.680696 | 84.000000 | 100.000000 | 0.893617 |
| A-D | 90.000000 | 100.000000 | 71.428571 | 86.666667 | 100.000000 | 0.928571 |
| A-E | 100.000000 | 100.000000 | 100.000000 | 100.000000 | 100.000000 | 1.000000 |
| B-C | 95.000000 | 95.454545 | 100.000000 | 100.000000 | 95.454545 | 0.954545 |
| B-D | 92.500000 | 100.000000 | 100.000000 | 100.000000 | 100.000000 | 0.938776 |
| B-E | 100.000000 | 100.000000 | 100.000000 | 100.000000 | 100.000000 | 1.000000 |

**Table 4.** Results of proposed model with NB Algorithm.

Tables 2, 3 and 4 represents the different performance measures for the K-NN, SVM and NB classifiers, respectively. The tables highlight the relative similarity between the different performance measures. From, the tables, it is observed that the highest accuracy i.e., a value of 100% was observed in the case of both A-E and B-E dataset. The recall, accuracy, and F-measure values achieved in the case of both the datasets A-E and B-E was 100%. However, for the rest of the datasets, F-measure values range since 0.950 to 0.996, the recall values range since 95% to 99% then the accuracy value range from 98.50% to 99.50%. The K-NN classifier achieved an accuracy 100% for the datasets: A-E, and B-E. Sensitivity and recall values of 100% was achieved for the datasets A-C, A-E, and B-E respectively. The lowest recall and sensitivity value of 95.3% was observed in case of the dataset A-C on using the K-NN classifier. Precision and recall values of 100% was observed in case of the datasets: A-C, A-D, A-E, B-D, and B-E on using the NB classifier. But the sensitivity and recall value of 95.45% was observed in the B-C dataset on using the NB classifier.

On using the A5 coefficients, the SVM classifier achieved a classification accuracy within 99.00 % accuracy and the K-NN classifier achieved a classification accuracy between 98.50 %. On using the detailed D5 coefficients, an accuracy of 99% and 98.50 % was observed in case of the SVM and the K-NN classifier respectively, and has also been represented in Tables 2 and 3. The Native Bayes classifier also achieved an accuracy above 95 %. It can be concluded that the values of different measures as well as precision, sensitivity, specificity, recall and F-measure were significant.

It can therefore be concluded that the relative discrete wavelet within frequency bands (1–3 Hz) and (4–6 Hz) was significant in classifying the eyes open and the eyes closed states from the EEG brain patterns.

## 4. Conclusion

In present paper, a novel model to detect the seizure (normal/seizure) from the EEG has been proposed. The 5th level of DWT has been used to substance the features since raw data after which, seven various statistical operations were applied on those extracted features. Finally, three machine learning methodologies were utilized for the classification of the signals. The proposed approach for the purpose of classification of epilepsy cases from the brain signals provide highly accurate result in comparison with relevant research. In the present work, we elaborate the significance of features extraction and features selection to generate efficiently performed outcomes, as a pre-processing step, before the EEG are classified. Accordingly, DWT were clubbed to witness effective and clear results for epilepsy diagnosis. Overall better classification accuracy has been gained, and it can be observed that the proposed algorithm can find a best subset of input features since the different DWT coefficients of the EEG signals. Performance of the classifiers has been evaluated in terms of precision, accuracy, and recall. From the experimental outcomes, it is observed that the SVM was the most efficient for the purpose of classification. Outcome also demonstrated this is NB and KNN produced near similar outcomes. SVM classifier resulted in the highest F-measure and Accuracy for the datasets. Future works may focus on testing the model for a larger EEG dataset, and DWT will also be utilized in order to attain more satisfactory outcomes.